\documentclass[aps,pra,preprint,groupedaddress]{revtex4}
\usepackage{color}
\usepackage{epsf}
\usepackage{epsfig}
\usepackage{graphicx}
\usepackage{latexsym}
\usepackage{bm}
\usepackage{amssymb}

\begin{document}

\title{Transfer excitation reactions in fast proton-helium collisions}

\author{M.~S.~Sch\"{o}ffler$^{1}$}
\email{schoeffler@atom.uni-frankfurt.de}
\author{H.-K.~Kim$^{1}$}
\author{O.~Chuluunbaatar$^{2,3}$}
\author{S. Houamer$^4$}
\author{A.G. Galstyan$^5$}
\author{J.~N.~Titze$^1$}
\author{T.~Jahnke$^1$}
\author{L.~Ph.~H.~Schmidt$^1$}
\author{H.~Schmidt-B\"{o}cking$^1$}
\author{R.~D\"{o}rner$^1$}
\author{Yu.~V.~Popov$^6$}
\author{A. A. Bulychev$^2$}

\affiliation{$^1$ Institut f\"ur Kernphysik, Universit\"{a}t Frankfurt, Max-von-Laue-Str. 1, 60438 Frankfurt, Germany}

\affiliation{$^2$Joint Institute for Nuclear Research, Dubna, Moscow region 141980, Russia}

\affiliation{$^3$School of Mathematics and Computer Science, National University of Mongolia, Ulaanaatar, Mongolia}

\affiliation{$^4$D\'{e}partement de physique, Facult\'{e} des Sciences, Universit\'{e} Ferhat Abbas, S\'{e}tif, 19000, Algeria}

\affiliation{$^5$Faculty of Physics, Lomonosov Moscow State University, Moscow 119991, Russia}

\affiliation{$^6$Skobeltsyn Institute of Nuclear Physics, Lomonosov Moscow State University, Moscow 119991, Russia}

\date{\today}

\begin{abstract}
Continuing previous work, we have measured the projectile
scattering-angle dependency for transfer excitation of fast
protons (300-1200 keV/u) colliding with helium (p+He $\rightarrow$
H + He$^{+ *}$). Our high-resolution fully differential data are
accompanied by calculations, performed in the plane-wave first
Born approximation and the eikonal wave Born approximation.
Experimentally, we find a deep minimum in the differential cross
section around 0.5 $mrad$. The comparison with our calculations
shows that describing the scattering-angle dependence of transfer
excitation in fast collisions requires us to go beyond the first Born
approximation and in addition to use the initial state-wave function,
which contains some degree of angular correlations.
\end{abstract}


\maketitle

\section{Introduction}

From an experimental point of view, single-electron transfer has
at least two interesting facets. First, it can be used as a tool
for spectroscopy \cite{fischer2002jpb}. Energy gain spectroscopy
and the related experiments in inverse kinematics exploit the
recoil ion longitudinal momentum for $Q$-value determination
(change in the electron binding energies). This allows us to
investigate the energy levels of highly charged species and/or
energy levels that do not decay radiatively or have long
lifetimes. Second, the dynamics of the transfer process itself is
of fundamental interest since it combines electron-electron
dynamics, correlation, and questions of few-body momentum
exchange.

While the ground-state charge transfer (CT), where no excitation
is involved, hardly allows us to access these interesting questions,
processes with two active electrons are much richer. Transfer
ionization, where one electron is captured into a bound state of
the projectile and a second one is released into the continuum, have
been studied in great detail \cite{Horsdal1986prl,Bossler1987pra,Palinkas1989prl,Shingal1991jpb,Mergel1997prl,Woitzke1998pra,Mergel2001prl,Schmidt2002prl,Schoeffler2005jpb,Godunov2005pra,Schulz2012prl,Schoeffler2013pra1,Schoeffler2013pra2}
and have led to a better understanding of initial and final state
correlation \cite{Shi2002prl,Schmidtb2003epl,Godunov2004jpb}.
Instead of being lifted to the continuum, the second target electron
can also be promoted to a bound but excited state. This so-called
transfer excitation (TE) has the experimental beauty that no
electron needs to be detected, which can be extremely challenging
especially at higher electron energies. An additional
projectile-electron interaction or a shake-up lifts the electron
into the excited state \cite{dunseath1991jpb,schoeffler2009pra}.

In most of the theoretical and experimental studies the transfer
into an excited state of the projectile or the transfer combined
with an additional target excitation of a second electron was
neglected. Especially at higher impact energies $E_p>100$ keV/u,
where the final electronic state determination in experiments is
challenging or often impossible, the influence of excitation has
not been investigated \cite{Horsdal1983prl,
fischer2006pra,fischer2010pra, fischer2010jpb}. Projectile
scattering angle distributions, which are final state selective,
are rather rare
\cite{mergel1995prl,doerner1998pra,abdallah1998pra,abdallah1998pra2,kamber1999pra,zhang2001pra,fischer2002jpb,knoop2008jpb,
schoeffler2009pra2}. All these measurements were only possible due
to the development of a modern momentum imaging technique, cold target recoil ion momentum spectroscopy (COLTRIMS) combined with
optimized three-dimensional focusing ion optics, which achieves
longitudinal momentum resolution $<$0.04 a.u. (FWHM)
\cite{kim2012pra}. This allows us to separate the final electronic
states even for p-He collisions at energies as high as 1.2 MeV. In
a previous publication we applied this high-resolution COLTRIMS
and investigated the pure electron transfer, CT (i. e., without
any excitation) \cite{kim2012pra}, while in the present paper we
present scattering-angle-dependent cross sections for transfer
excitation (i. e. the remaining target electron became excited).

The general theory of ion-atom collisions has been formulated for
a long time by many authors, and one can find rather full reviews
of the literature in \cite{Belkic79, Belkic08, Belkic_book}. We
recall that the plane-wave first Born approximation (PWFBA)
is valid only if the asymptotic wave functions in both the
entrance and exit channels satisfy Dollard's condition
\cite{Dollard}. Fortunately, in the case of the
particular reaction considered, p+He $\rightarrow$ H + He$^{+ *}$, these
conditions are fulfilled. In the initial asymptotic state the
charged proton does not interact with the neutral helium atom by
long-range Coulomb force, and in the final asymptotic state
the neutral hydrogen atom does not interact with the charged residual
ion He$^{+*}$. In this case PWFBA $\equiv$ CB1 (corrected first
Born approximation) \cite{Belkic08}.

Recently, Madison and colleagues published a series of papers with
calculations of the fully differential cross sections (FDCS) for
the transfer excitation reaction considered here at various
energies \cite{Madison_tot}. A code for nine-dimensional (9D) numerical integration
was used in this study. The results differed from the experiment
by the factor $v_p^4$. Some of the present authors showed in a comment
\cite{Salim_com} that this disagreement can be attributed to the quality
of the numerical code
\footnote{Recently D. Madison
communicated that they found another reason for such a
disagreement.}. We will add a more detailed discussion and new 9D
results in Sec. III B below.

Here we consider the transfer excitation reaction p + He $\to$
H + He$^{+*}$ at different high proton energies (300 - 1200 keV) and
present both the experimental single differential cross sections
for total excitation of the residual helium ion ($n\geq 2$) and
the calculations within the PWFBA and the eikonal wave Born approximation (EWBA). Atomic
units $\hslash = e = m_e = 1$ are used throughout unless otherwise
specified.

\section{Experiment}

At the high impact velocities (3.5 - 7 a.u.) investigated here,
the best energy loss and scattering-angle resolution is obtained
by detecting the recoil-ion momentum instead of the momentum
change of the projectile \cite{ali1992prl,mergel1995prl}. This
corresponds to a transformation from the projectile to the
laboratory frame. In the present experiments we used the well-established COLTRIMS technique to measure both the neutral
projectile H and the recoiling He$^+$ ion in coincidence
\cite{Ullrich1997jpb,Doerner2000pr,Ullrich2003rpp}. The experiment
has been performed at the 2.5 MV Van de Graaff accelerator at the
Institut f\"ur Kernphysik, Universit\"{a}t of Frankfurt. The proton
beam was collimated with two sets of adjustable slits to a size of
0.5$\times$0.5~mm$^2$ at the interaction point. An electrostatic
deflector placed upstreams of the target was used to remove beam
impurities (H) by deflecting the primary beam slightly upwards.
Behind the interaction region, right after leaving the
spectrometer, another electrostatic deflector horizontally
separated the primary beam of protons from the charge exchanged H.
While the main proton beam ($\approx 1~nA$) was dumped in a 0.5 m
long Faraday cup, the neutral H projectiles were detected on a
40~mm position- and time-sensitive microchannel plate (MCP)
detector with a delay line anode 
\cite{Jagutzki2002nima1,Jagutzki2002nima2} 5 m downstream of the target region. The target was
provided by a two-stage supersonic gas jet of helium atoms. At the
interaction point, the gas jet had a diameter of $\approx 1.5~mm$
and an area density of 2$\times10^{11}$ atoms/cm$^2$. The He$^+$
recoil ions produced in the overlap region of the gas jet and
projectile beam were projected by a weak electrostatic field (9
V/cm) onto an 80-mm position- and time-sensitive MCP detector. To
achieve the best possible momentum resolution, a three-dimensional
time- and space focusing geometry was applied
\cite{doerner1997nimb, schoeffler2009pra, kim2012pra}. From the
measured data, the time of flight (16~$\mu$s for He$^+$), and the
position of impact, the initial three-dimensional momentum vector
of the recoiling ion was derived. A momentum resolution $<$0.04
a.u. in the longitudinal direction ($p_{||}$), in which the
$Q$ value of the reaction is encoded [see \cite{ali1992prl,
mergel1995prl}], was achieved (see Fig. 2(c) in
\cite{kim2012pra}). The spectrometer geometry and electric fields
yielded a 4$\pi$ acceptance angle for all He$^+$ ions with momenta
below 9 a.u.

In the plane perpendicular to the beam axis, we measured the
scattering angle of the projectile and the transverse momentum of
the recoiling ion. By momentum conservation they must add to zero, which was
used for background suppression. The scattering angles
presented below were deduced from the He$^+$ transverse momentum
$p_{\perp}$, which has a far better momentum resolution ($<$0.1
a.u.) than the projectile. Gates on the different longitudinal
momenta of the recoil ion ($p_{||}$) allow us to extract the
scattering-angle distribution for a certain final electronic state
\cite{schoeffler2009pra2}. The  remaining tiny background of
statistically false coincidences has been subtracted.

The data presented here (similar to \cite{kim2012pra}) show a
significantly different scattering angle distribution than
\cite{fischer2010pra}. A notorious problem in most COLTRIMS
measurements which might lead to differences in the results from
different works is the calibration of the momentum. We therefore
describe our calibration procedure in some detail to highlight its
reliability on a 1~\% level. We choose our 1200 keV data set for
this purpose. The position to momentum conversion along the beam
direction (z) is based on the $Q$-$p_{||}$ relation
(\cite{ali1992prl, mergel1995prl}). As the $Q$ value and the beam
velocity are known on an absolute scale, the momentum can be
calculated directly (see Fig. 2 in \cite{kim2012pra}). For the
charge transfer reactions investigated here, the longitudinal
momenta for the various final electronic states (no excitation or 
target or projectile or both being excited) differ in total by less
than 0.5 a.u. Background events originate from a single ionization
peak at $p_{||}=0$ \cite{doerner95jpb} and are therefore an
excellent cross-check for our calibration on a larger scale. As
can be seen from Fig. 1(a), the measured and calculated
longitudinal momenta are in excellent agreement. A linear fit
through these data points confirms the high quality of this
calibration with a deviation of less than 1 \%. This momentum is
measured via the horizontal position of impact on the ion
detector.

In the direction of the gas jet (y, vertical) the same position to
momentum calibration factor was applied, utilizing the cylindrical
symmetry of our spectrometer. The momentum in the time-of-flight
(x) direction depends linearly on the electric field, which is
known quite accurate and is additionally compared with SIMION
simulations. Furthermore the physical symmetry around the beamaxis
was checked by comparing the momenta $p_x$ (time-of-flight
direction) and $p_y$ (position direction). Figure 1(b) shows the
transversal recoil ion momentum $p_\perp$ vs. the corresponding
azimuthal angle $\phi$ around the beam axis $p_x$/$p_y$. The
physical symmetry is well reproduced, confirming the consistency
of our momentum calibration in time-of-flight and position
directions of our spectrometer.

\section{Theory}

\subsection{General formulas}

Let us denote the projectile proton momentum by $\vec p_p$, the
hydrogen momentum by $\vec p_H$, and the recoil-ion momentum by
$\vec K$. We also define the transferred momentum as $\vec q=\vec
p_H-\vec p_p$. The proton mass is $m_p=1836.15$, the helium ion mass is $M\approx
4m_p$, and the ground-state energy $E_0^{He} = -2.903724377034$.

We choose very small scattering angles for the outgoing hydrogen
($0\leq\theta_{p,lab}\lesssim 1.5$ $mrad$). It leads to a
practically zero ion velocity $K/M$ in the laboratory frame, and
we can consider the ion as immovable during the reaction. The
proton velocity $v_p=p_p/m_p$ and $q$ vary about a few atomic units for
a proton energy of several hundred keV. This fact allows us to
neglect ${K^2}/2M$ and $q^2/2m_p$ everywhere they appear. Choosing
the vector $\vec v_p$ as the $z$ axis, we obtain from the energy
and momentum conservation for the longitudinal component
$q_z={v_p}/{2}+{Q_n}/{v_p}$. The transverse component is
$q_\perp=(\vec p_H)_\perp\approx m_pv_p\ \theta_{p,lab}$, and
$Q_n=E_0^{He}-E^H_0-E^{ion}_n.$

The single differential cross section (SDCS) for TE processes takes the form
\begin{equation}
\frac{d\sigma_{ex}}{d\theta_{p,lab}} = 2\frac{m_p^2
\theta_{p,lab}}{(2\pi)}\
\sum_{n=2}\sum_{l=0}^{n-1}\sum_{m=-l}^{l}\vert T_{nlm}\vert^2
 \end{equation}
The factor 2 in (1) appears as a result of symmetrization of the final
wave function.

The nonsymmetrized  first Born approximation (FBA) matrix element in (1) is presented by the
well-known 9D integral (see, for example, \cite{JS})
$$
T^{FBA}_{nlm}\approx \int d^3R e^{-i\vec R\vec q}\int d^3\rho
e^{i\vec\rho\vec v_p}\ \varphi_0(\rho)
$$
\begin{equation}
\int d^3 r_2 \phi_{nl}(r_2)Y_{lm}(\vec r_2)\
\left[-\frac{1}{\rho}-\frac{1}{|\vec R-\vec
r_2|}+\frac{2}{R}\right]{\Phi}_0(\vec R-\vec\rho,\vec r_2).
\end{equation}
The vector $\vec R$ is the center of mass of the moving hydrogen
subsystem at the end of reaction, $\vec\rho$ is the relative
coordinate of electron 1 (transferred) in the hydrogen
[described by the ground wave function $\varphi_0(\rho)$], and
$\vec r_2$ is the relative coordinate of electron 2 in the
He$^+$ subsystem (see details in \cite{Salim10,kim2012pra}). The
hydrogen like excited (ground) wave function of the residual ion
He$^+$ $\phi_{nl}(r)Y_{lm}(\vec r)$ can be found in any text book.

Four different trial ground-state helium wave functions were used
for the calculations. One is the loosely correlated $1s^2$
Roothaan-Hartree-Fock (RHF) wave function \cite{RHF} (no angular
correlation) with a rather poor ground-state energy of
$E^{RHF}_{He}=-2.8617$ a.u. The second one [Silverman, Platas, and
Matsen (SPM) \cite{SPM}] includes angular correlations, but its
ground energy $E^{SPM}_{He}=-2.8952$ is also far from the
literature value. Two other trial functions are highly correlated. They are given in \cite{Chuka} with a ground-state energy of
$E^{Ch}_{He}=-2.903721$ a.u. and \cite{Mitroy}, with a ground-state energy of $E^{Mitroy}_{He}=-2.9031$. Their energies are very close to the best-known
ground-state energy. After the Fourier transformation of the wave
functions in (2) we come to three-dimensional (ED) integrals for configuration interaction helium wave
functions \cite{RHF,SPM,Mitroy} and four-dimensional (4D) integrals for the function
\cite{Chuka}.

\subsection{9D integration}

In the case of the SPM helium wave function we can also calculate
EWBA matrix elements using the
follwoing code for 9D integration:
$$
T^{EWBA}_{nlm}\approx \sqrt{2}\int
d^3R e^{-i\vec R\vec q}\int d^3\rho e^{i\vec\rho\vec v_p}\
\varphi_0(\rho)
$$
\begin{equation}
\times \int d^3 r_2 \phi_{nl}(r_2)Y_{lm}(\vec r_2)\ e^{-(i/v_p) \ f(\vec
R,\vec\rho,\vec r_2)}\left[-\frac{1}{\rho}-\frac{1}{|\vec R-\vec
r_2|}+\frac{2}{R}\right]{\Phi}_0(\vec R-\vec\rho,\vec r_2),
\end{equation}
with the eikonal phase factor
\begin{equation}
f(\vec R,\vec\rho,\vec r_2)=\ln\left[\frac{[v_p|\vec
R-\vec\rho|+\vec v_p\cdot(\vec R-\vec\rho)]^2\ [v_p|\vec R-\vec
r_2|+\vec v_p\cdot(\vec R-\vec r_2)]}{[v_pR+\vec v_p\cdot\vec
R]^2\ [v_p|\vec R-\vec\rho-\vec r_2|+\vec v_p\cdot(\vec
R-\vec\rho-\vec r_2)]}\right].
\end{equation}
The way to get this phase factor can be found in \cite{kim2012pra}.

Here we have to say a few words about 9D integration of
oscillating functions. Chowdhury et al. recently presented a
series of papers (see, for example, \cite{Madison}) where they
calculated 9D integrals the types in (2) and (3) for TE reactions.
They obtained a discrepancy of a factor of 150 between theory and
experiment at $E_p= 300$ keV. In \cite{Salim_com,Madison_rep} some 
of the present authors have attributed this to a numerical problem
of a non-optimal code of successive integration.

In our code we use the so-called adaptive subdivision method
(ASM). Both the open Fortran codes \cite{adapt,dcuhre} and commercial
ones \cite{nag} are available and written on the basis of the
adaptive subdivision method. We modified these codes, and now they
keep more data in the memory, can use the complex arithmetics, and
are adapted for parallel calculations.

For the oscillating functions in (2) and (3)  hyperspherical
variables are used. In this case the integral by hyperradius is
calculated analytically. The remaining 8D integral does not
contain oscillations, and its domain of integration is restricted.
For $E_p=300$ keV the 8D integral is calculated with the relative
accuracy $\epsilon=0.1$. To approach this accuracy for
$\theta_{p,lab}=1$ $mrad$, it takes about $10^8$ subregions of the
ASM; 401 points of integration are taken in each subregion
(so-called 7-point rule), i.e., about 21 points per variable.

\subsection{Estimate of the residue in (1)}

We note that the value $Q_n$ satisfies an inequality $-2.403 <
Q_n\leq -0.403$ for any $n\geq 1$. If we apply the closure
approximation, i.e. replace $Q_n\to \bar Q$  in the sum by all
bound states with $-2.403 < {\bar Q}< -0.403$, we obtain, using the
completeness condition of the Coulomb spectral functions,
\begin{equation}
\sum_{n=1}^\infty\sum_{l=0}^{n-1}\sum_{m=-l}^{l}\ |T_{nlm}|^2 =
\int\frac{d^3 k}{(2\pi)^3}|<\vec k|T>|^2 - \int\frac{d^3
k}{(2\pi)^3}|<\varphi^-(\vec k)|T>|^2.
\end{equation}
The symbol $<\varphi^-(\vec k)|T>$ denotes the transfer ionization
amplitude, where the emitted electron is described by a Coulomb
wave \cite{Schoeffler2013pra1,Schoeffler2013pra2}; $<\vec k|T>$ is
the same, but the Coulomb wave is replaced by the plane wave. Here
we have 6D and 7D integrals. We recall that in all integrals in
(5) only $\bar Q$ is used. Below we show that the presentation (5)
allows us to estimate the residue of the sum in (1) beyond the exact
calculations with a given $n$.

\section{Results and discussion}

In Fig. 2 experimental and theoretical single-differential cross
sections ($d\sigma/d\theta_{p,lab}$) for TE within the first Born
approximation for various initial states are presented.
Experimentally we can no distinguish between the different excited
states; e.g., we sum over all $n\geq2$. The integral of the
electron transfer (with and without excitation) has been
normalized to that in \cite{williams1967pr,shah1989jpb}. TE contributes
4$\%$-5$\%$ to the total electron transfer cross section, while the
CT without any additional excitation amounts to
75$\%$-80$\%$ \cite{kim2012pra}. The calculated and measured total
cross sections (TCS) for TE, shown in Table I, are in good
agreement only for the well correlated helium wave functions
\cite{Chuka,Mitroy}. The calculation based on a wave function
without angular correlation completely fails. Obviously, TCS are
defined mainly by the angle integration around the main peak in
Fig. 2, where theory is close to the experiment for correlated
helium wave functions. Comparing calculated and measured TCS has
been the benchmark for testing theories. However, this is
necessary but not sufficient.

\begin{table}[htb]
  \begin{center}
    \begin{tabular}{|r|c|c|c|c|c|}
    \hline
      & Experiment & Ref. \cite{Chuka} & RHF & \cite{Mitroy} & Ref. SPM \\
      & (black dots) & (black solid line) & (red dashed line) & (green dotted line) & (blue dash-dotted line) \\ \hline
    300 keV & 1.093 a.u. & 1.137 a.u. & 0.8593 a.u. & 1.150 a.u. & 1.204 a.u. \\
    630 keV & 0.0404 a.u. & 0.0581 a.u. & 0.0204 a.u. & 0.0581 a.u. & 0.0616 a.u. \\
    1000 keV & $8.39\times10^{-3}$ a.u. & $7.08\times10^{-3}$ a.u. & $1.41\times10^{-3}$ a.u. & $7.16\times10^{-3}$ a.u. & $7.76\times10^{-3}$ a.u. \\
    1200 keV & $2.79\times10^{-3}$ a.u. & $2.93\times10^{-3}$ a.u. & $4.79\times10^{-4}$ a.u. & $2.98\times10^{-3}$ a.u. & $3.29\times10^{-3}$ a.u. \\
    \hline
    \end{tabular}
    \caption{Total transfer excitation cross section for p+He collisions at 300-1200 keV impact energy.
    Experimental data have been normalized to published total electron transfer cross
    sections \cite{williams1967pr, shah1989jpb}. Theoretical data are shown for various wave functions.}
    \label{table1}
 \end{center}
\end{table}

The experimental distribution is overall similar to the one for CT
without excitation \cite{kim2012pra} and exhibits two
well-pronounced domains. Around 0.1 $mrad$ a sharp peak with a
steep decrease is followed by a rather flat contribution
for $\theta_{p,lab}>0.5$ $mrad$. The peak at small
angles originates from the momentum kick of the transferred
electron \cite{Kamber1988prl,doerner1989prl,Mergel1997prl}. If the
captured electron is assumed to be at rest, the maximum possible
proton scattering is 0.55 mrad; larger scattering angles
require a momentum transfer between the nuclei. Especially, the minimum between these two contributions is
more pronounced for TE than it is for CT. Also, the minimum is
shifted slightly towards larger scattering angles
($\theta_{p,lab}=0.45$ $mrad$ for CT, $\theta_{p,lab}=0.5$ $mrad$
for TE at 1.2 MeV). These findings contradict the
measurements reported by Fischer et al. \cite{fischer2010pra}, who
found the minimum at $\theta_{p,lab}=0.35$ $mrad$ for 1.3 MeV and
have a narrower distribution in general. To exclude the most
obvious possible source of error we have cross-checked that the
calibration of our momenta is accurate within 1\%, as described
in Sec. II. 

The minimum shows the boundary angle between the two main capture mechanisms. The shake-off amplitude ($A_1$+$A_3$); defined and explained in \cite{Schoeffler2013pra1}) strongly depends on the electron-elctron-correlations, defines the position and amplitude of the main peak, but goes down quickly. The sequential amplitude ($A_2$) is partially represented by the FBA but requires higher Born terms to be taken into account. The minimum defines some boundary between them. In the case of CT this minimum is sharp and equal zero for FBA. For TE it is not so sharp and is washed out by the excitation. But the minimum's origin remains the same. So we see that FBA (EWBA) can influence the amplitude and position of the main peak but not the position of the minimum. Calculations of higher Born terms are needed to account also for the sequential mechanism.

The experimental and the theoretical minima do not seem to be connected, especially as the experimental one shifts towards smaller angles with increasing projectile velocity. Therefore we speculate that the experimentally observed dip around 0.5 mrad can be explained in the following way. The broad contribution, peaking at larger scattering angles (1.5 mrad at 630 keV and 1.0 mrad at 1200 keV), may well be a result of the well-known Thomas-process (for details see Kim et al. \cite{kim2012pra} and references therein). This classical double-scattering process ideally predicts a sharp peak at 0.47 mrad and was found to be around this value for very high energies \cite{Horsdal1983prl,fischer2006pra,fischer2010pra}. At the rather low energies presented here, this peak might be shifted towards larger $\theta_{p,lab}$ due to an additional nucleus-nucleus scattering (N-N). From the strict geometrical conditions leading to Thomas-like electron capture, the momentum kick from the N-N scattering always points in the same direction as the initial kick with the electron. Hence the overall transverse momentum of the projectile $\theta_{p,lab}$ is the largest at small $v_P$, decreases for faster projectiles, and, finally, converges to the value of 0.47 mrad at an infinite projectile velocity. 

We limit our calculations to $n\leq 3$ in the sum [Eq. (1)]. The shape of the SDCS is formed by three terms in Eq. (2), one of which (OBK) provides the direct He$\to$ e+He$^+$ decay mechanism; the other two provide the double decay He$\to$
2e+He$^{2+}$ in the intermediate state. The shape in the case of
helium wave functions with angular correlations (all except the
red dashed line in Fig. 2) shows a the finite minimum very similar to the
case $n=1$ \cite{kim2012pra}. The wave function without angular
correlations (red dashed line) fails completely. It is interesting
to note that the SDCS calculated within the FBA for the CT
practically does not depend on the trial helium wave function, and
even the simplest $1s^2$ wave function describes the main peak
\cite{kim2012pra} for the one-electron process quite well. The
calculated total cross sections with angular correlated initial-state wave functions agree well with the experiment, as can be
seen in Table I.

We also see that $E_p\sim 500$ keV is the lowest energy for which FBA
still somehow describes the main peak at very small
$\theta_{p,lab}$. For $E_p= 300$ keV the FBA fails completely. The
FBA also fails to describe the position of the minimum and the
behavior of the cross section beyond it. This is again in
agreement with the conclusion drawn from CT (no excitation) in
\cite{kim2012pra}. Calculations in the distorted-wave Born approximation or
those performed in the second Born approximation are needed to
describe the angular scattering in the regime where both the
momentum transfer from the captured electron and the momentum
exchange between the nuclei contribute.

We have tried to improve the description of the scattering by
including an eikonal phase factor [Eq. (4)]. The results of this EWBA
calculation for $E_p=$630, 1000, and 1200 keV are presented in
Figs. 3(a)-3(c). We use 9D integrals [Eq. (3)] with the SPM trial
helium wave function and the eikonal phase factor (4). The positive
(but very slow) progression of the main peak about 0.1 mrad
towards the experiment is clearly visible. However, this EWBA can
not improve the situation at larger scattering angles; full second
Born calculations are needed. We currently cannot perform these
calculation for TE. For CT, however, second Born calculation are
feasible. To illustrate the influence of the second Born approximation (SBA) on the
angular scattering we present in Fig. 4 previously published
\cite{kim2012pra} calculations in the first and second Born
approximations together with EWBA calculations [Eq. (4)] for the CT
reaction at 630 keV when the residual He$^+$ ion stays in its
ground state ($n=1$). While the EWBA considerably improves the
agreement, the SBA calculations nearly perfectly describe the
experimental data. Of course, the 9D numerical results fully
coincide with 3D calculations of [Eq. (2)] and are close to the
experiment, at least in its absolute scale.

In Fig. 5 the results of calculations with the approximation from
[Eq. (10)] compared with exact calculations
$n\leq 3$ are shown for two
different wave functions [RHF in Figs. 5(a) and 5(c), and SPM in Figs. 5(b) and 5(d)].
In spite of rich behavior of the SDCS at very small scattering
angles for different fitting parameters $\bar Q$, all curves merge
into one bundle beyond some projectile-energy-dependent scattering
angle. For $E_p=1200$ keV it is about 0.6 $mrad$. That bundle has a different amplitude than the exact curve, but has a similar shape. This difference
allows us to estimate a residue of the sum over exited states, for $n>3$ to be about 10~\% for the loosely correlated RHF wave
function [Fig. 5(a) and (c)]. With the SPM wave function, which
includes both radial and angular correalation in figure 5 (b) and
(d) we see practically no gap (about 1$\%$-2$\%$, and
$n\leq 3$ calculations are enough
to describe the angular spectrum).

In \cite{schoeffler2009pra} the dynamical process that leads to a
transfer excitation was investigated for collision energies 60-300
keV. The ratio of the transfer excitation and charge
transfer ($R=TE/(TE+CT)$) of the single differential cross section
$d\sigma/d\theta_{p,lab}$ was plotted. A peak around 0.4 $mrad$
was found in collisions of $p$ and $He^{2+}$ colliding with
helium. Similarly, in Fig. 6 the same ratio of transfer
excitation and charge transfer for higher impact energies up to
1.2 MeV is plotted. Here a clear peak around 0.4 $mrad$ can
also be observed, which shifts towards smaller scattering angles with
higher projectile energy. This peak arises because the
energy transfer necessary for the excitation goes along with a
momentum transfer from the projectile to the electron to be
excited, which is also in the transverse direction \cite{Horsdal1986prl}.
This shifts the scattering angle to larger values, approaching 0.55
$mrad$, the maximum scattering angle of a proton at an electron.
For large scattering angles the scattering is caused by the
momentum transfer between the nuclei, and hence CT and TE show a
similar fall off.

In Fig. 6(e) we present the ratio $R$ for EWBA and PWFBA
calculations when the helium function is highly correlated. The
angle domain $0<\theta_{p,lab}<0.4$ $mrad$ is the most interesting
for a comparison of theory and experiment in the case of our
simple approximations. Here we are still near the main peak 
for both TE and CT reactions. We see that the agreement with the experiment is rather poor if
PWFBA is used. However, EWBA describes the domain of interest much
better. It is a good sign that this approximation improves the
description of the SDCS in the vicinity of main peaks at high
impact energies.

\section{Conclusions}

In conclusion, we present experimental data of the cross-section
differential in the scattering angle and the corresponding FBA
theory for transfer excitation in proton-helium collision at 300,
630, 1000, and 1200 keV. Our calculations have been carried out
using the plane-wave first Born approximation with $1s^2$ and highly
correlated trial helium wave functions. We find that this two-electron process cannot be described using a $1s^2$ wave function.
All wave functions including angular correlations give rather
similar scattering-angle distributions. All fail to describe the
experiment for most scattering angles. Including an eikonal
factor does not solve this problem. Calculations for capture
without excitation show that a second Born theory describes the scattering correctly, at least for this channel. The projectile
scattering is determined by an interplay of the momentum exchange
of the projectile with the target nucleus, the electron which is
captured, and the electron which is excited. Thus a theory capable
of describing this process must include an angularly correlated
initial-state wave function as well as second- or higher-order Born
terms.

\section{Acknowledgments}

We acknowledge financial support from the Deutsche
Forschungsgemeinschaft (DFG), Grant No. SCHO 1210/2-1. A.G.G. is
grateful to the Dynasty Foundation for the partial financial
support. This work was also partially supported by the Russian
Foundation of Basic Research (RFBR), Grant No. 11-01-00523-a. All
calculations were performed at the Moscow State University Research
Computing Centre (supercomputers Lomonosov and Chebyshev) and
the Joint Institute for Nuclear Research Central Information and
Computer Complex. The authors are grateful to D\v{z}. Belki\'{c}
for inspiring discussions and help.

\bibliographystyle{unsrt}

\newpage

\begin{figure}[htb]
  \centering
    \includegraphics[width=6cm]{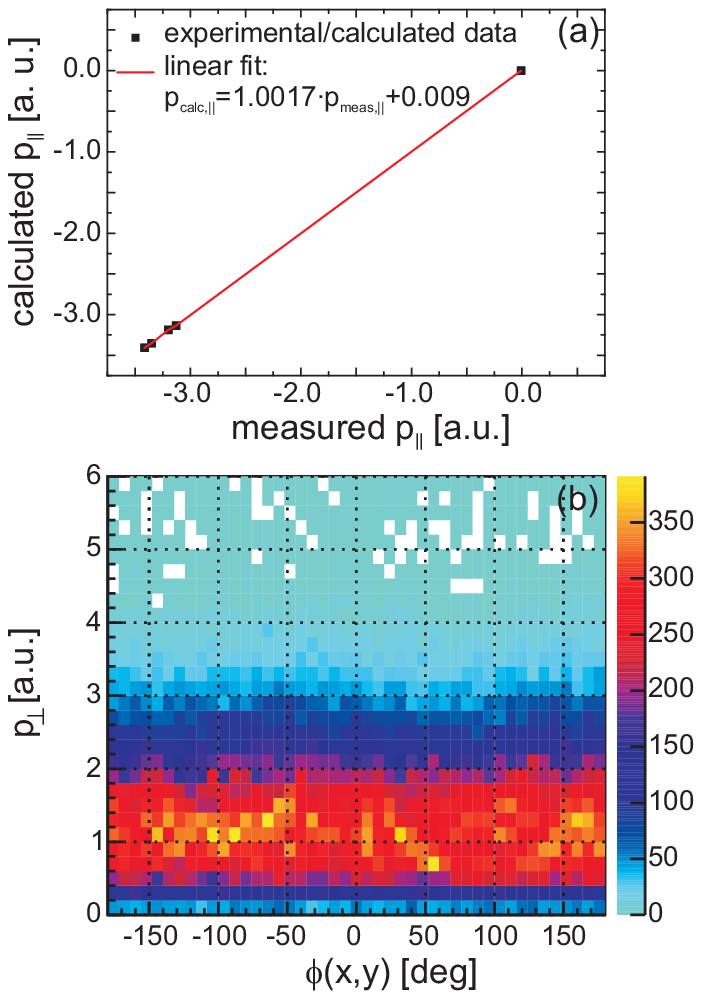}
      \caption{(Color online) Check of the experimental calibration for electron
      transfer in p+He collisions at 1200 keV. (a) Measured longitudinal ($p_{||}$)
      momentum of the distinguishable four final states for electron transfer and single
      ionization (peaking at zero). An additional linear fit through the data points shows the accuracy
      of the $p_{||}$ calibration. (b) $\phi$ angle around the beam axis vs.
      transversal momentum $p_\perp$, showing the proper calibration of the time-of-flight
      and position direction for electron transfer.}
\end{figure}

\newpage

\begin{figure}[htb]
  \centering
    \includegraphics[width=6cm]{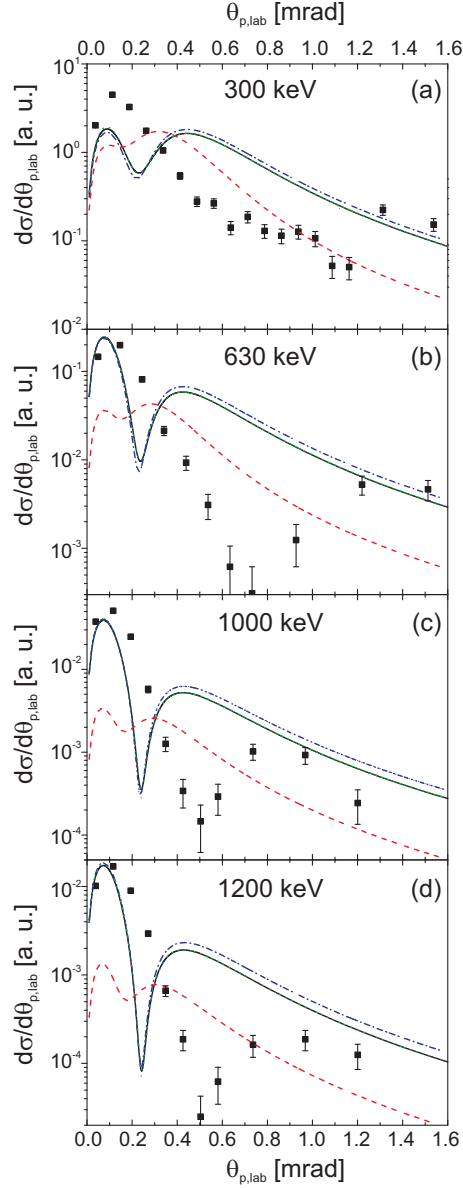}
      \caption{(Color online) Experimental and theoretical data for the scattering-angle-dependent transfer excitation in p+He collisions. Squares are the experimental
       points with statistical error bars, the dashed red line is the RHF \cite{RHF} trial helium
       wave function, the dash-dotted blue line is the  SPM \cite{SPM} wave function, the solid black line \cite{Chuka}
       and the dotted green line \cite{Mitroy} practically coincide. Experimental data have been
       normalized to published total electron transfer cross sections
       \cite{williams1967pr, shah1989jpb}. (a) 300 keV, (b) 630 keV, (c) 1000 keV and
       (d) 1200 keV incident proton energy. Experimental error bars show the statistical standard deviation.}
\end{figure}

\newpage

\begin{figure}[htb]
  \centering
    \includegraphics[width=6cm]{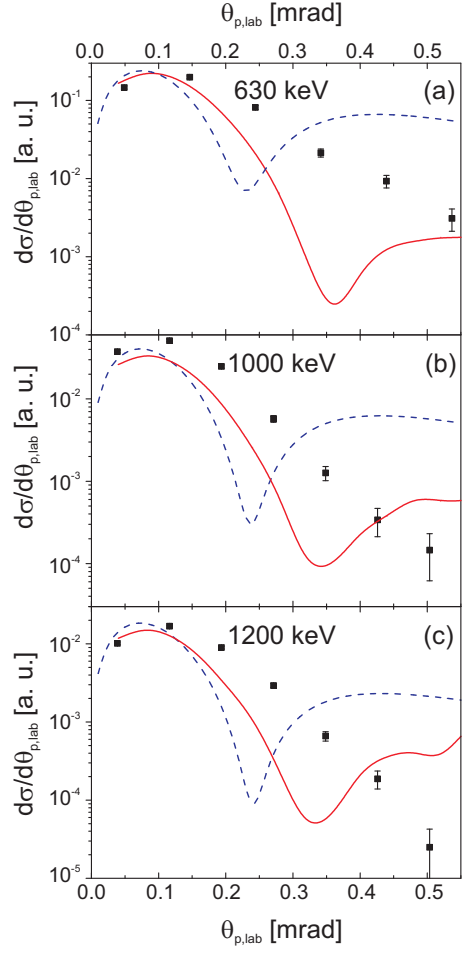}
    \caption{(Color online) Experimental and theoretical data for the
    scattering-angle-dependent transfer excitation in p+He collisions
    for (a) 630 keV, (b) 1000 keV, and (c) 1200 keV impact energy.
    SPM FBA calculations are supplemented by the eikonal phase factor.
    Dashed blue line: SPM; solid red line: SPM with the
    4C phase factor; full squares: experimental data (same as in Fig. 2). }
\end{figure}

\newpage

\begin{figure}[htb]
  \centering
    \includegraphics[width=6cm]{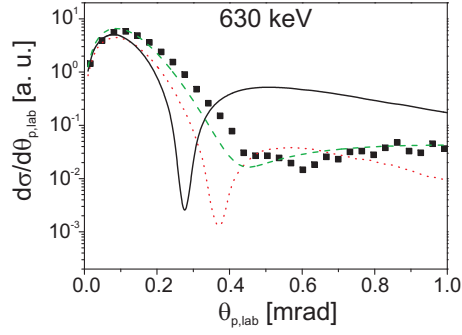}
    \caption{(Color online) SDCS vs. the scattering angle
    $\theta_{p,lab}$ for the CT process leaving the helium ion in its
    ground state $n=1$. The RHF helium wave function is used.
    Solid black line: FBA; dashed green line: SBA (taken from \cite{kim2012pra}, Fig. 3, $\bar E=0.1$);
    dotted red line: EWBA; squares: experiment \cite{kim2012pra}. $E_p=630$ keV.}
\end{figure}

\newpage

\begin{figure}[htb]
  \centering
    \includegraphics[width=12cm]{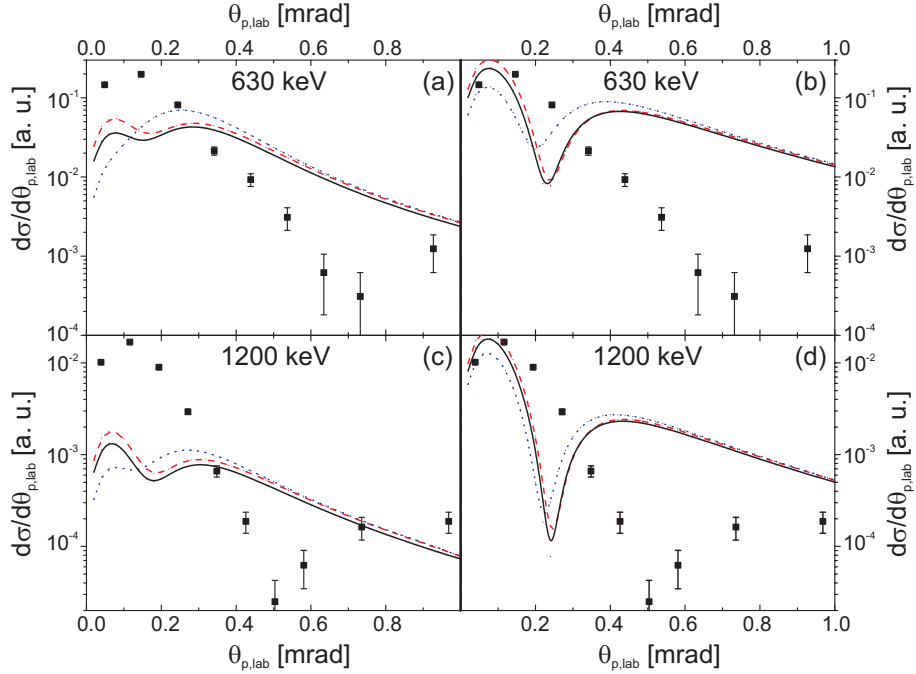}
    \caption{(Color online) SDCS vs.  the scattering angle $\theta_{p,lab}$
    for the TE process for (a) and (b) 630 keV and (c) and (d) 1200 keV.
    The RHF wave function is used for the calculations in (a) and (c),
    while the SPM wave function is used in (b) and (d).
    Solid black line: Eq. (5) with $n$=2+3; dotted blue line: Eq. (10)
    with $\bar Q$=-0.403 a.u.; dashed red line: Eq. (10) with $\bar Q$=-2.403 a.u.;
    squares: experimental data.}
\end{figure}

\newpage

\begin{figure}[htb]
  \centering
    \includegraphics[width=12cm]{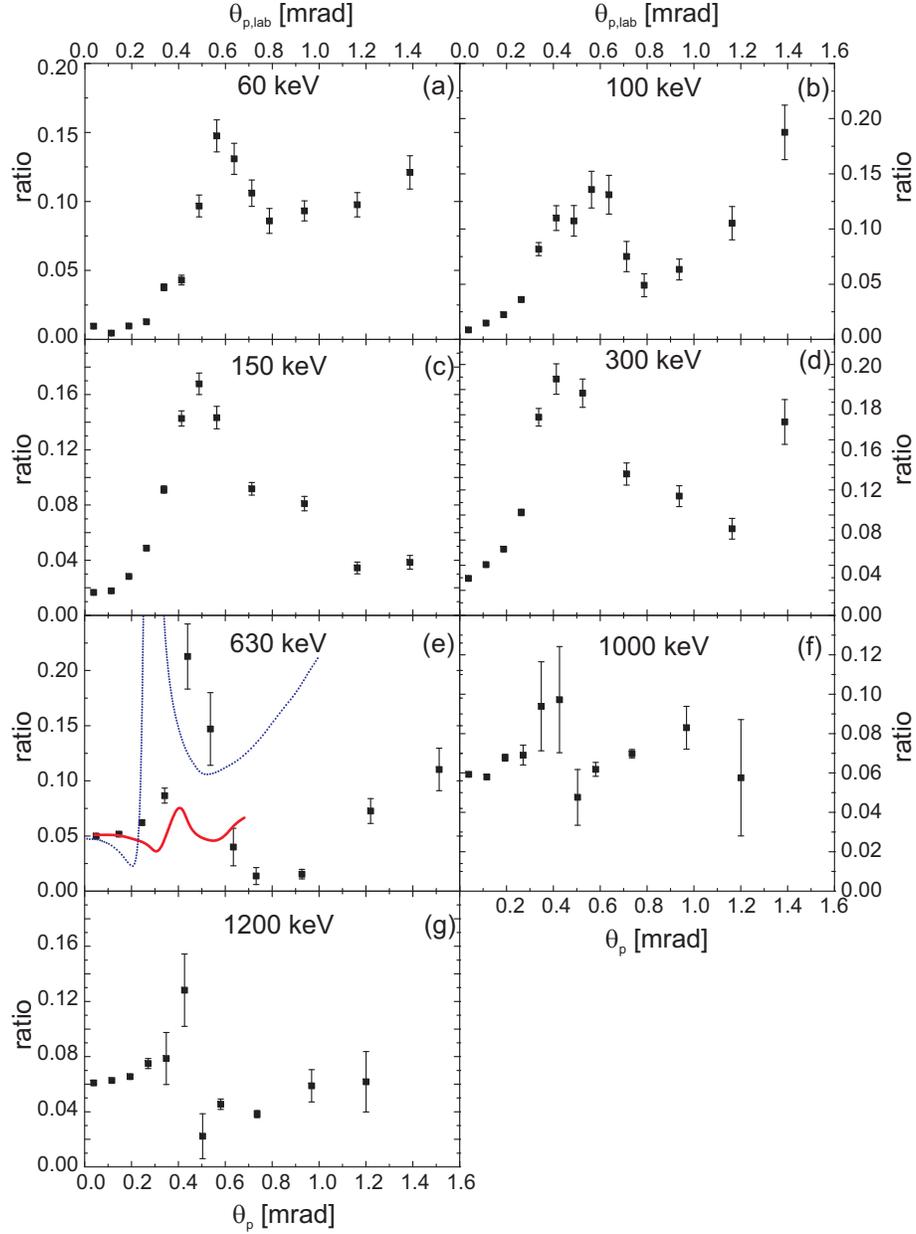}
    \caption{(Color online) Ratio of transfer excitation and charge transfer of the
    differential cross section $d\sigma/d\theta_{p,lab}$ in p+He collisions
    at projectile energies of 60-1200 keV. Datapoints for (a)-(d) were taken from \cite{schoeffler2009pra}.
    Black squares are the experimental points. The solid red line in (e) represents an EWBA and the dotted blue line represents a PWFBA calculation.}
\end{figure}

\end{document}